\documentclass[preprints,article,accept,moreauthors,pdftex]{Definitions/mdpi} 

\firstpage{1}
\pubvolume{1}
\issuenum{1}
\articlenumber{0}
\pubyear{2023}
\copyrightyear{2023}
\datereceived{}
\dateaccepted{}
\datepublished{}
\hreflink{https://doi.org/} 
\usepackage{xcolor}

\Title{Constraints on black hole charges in M87* and Sgr A* with the EHT observations}

\TitleCitation{Constraints on black hole charges in M87* and Sgr A* with the EHT observations}

\Author{Alexander F. Zakharov$^{1,*}$\orcidD{}}

\AuthorNames{ Alexander F. Zakharov}

\AuthorCitation{ Zakharov, A. F. }

\address{%
$^{1}$ \quad Bogoliubov Laboratory for Theoretical Physics, JINR, 141980 Dubna, Russia \\
}

\corres{Correspondence: alex.fed.zakharov@gmail.com}

\abstract{
In May 2022 ICRANet organized the Workshop dedicated to the 80th anniversary of Professor Ruffini.
This paper is based on the talk delivered at the meeting.
 Professor Ruffini was well known for Soviet scientific community not only due to his publications in leading journals
but also due Russian translations of his books where he was an author or a contributor in collection of articles.
 But only in 1988 I
 had an opportunity to watch and listen professor R. Ruffini at the Conference dedicated to the century since the birthday of Alexander Alexandrovich Friedmann. This conference was organized in Leningrad (Soviet Union) in June during a short magic period when there are white nights there.  In June 2023 we celebrate the 135th anniversary of Friedmann's birth.
Friedmann and his closed friend V. K. Frederics were the founders of Soviet school of general relativity and George Gamow was one of the brilliant representative of the school and he was the author of the hot Universe model which is the most popular now. In the USSR a development of general relativity and relativistic cosmology was not smooth and only in sixties of the last century these branches of science freed from the total control of representatives of the ideology of Marxism -- Leninism.
I also discussed a Soviet contribution in a discovery of cosmic microwave background radiation done by T. Shmaonov in 1957
and reasons why his supervisors did not connect these results with the hot Universe models discussed by G. Gamow.
Author's results about observational features of supemassive black holes (including the black hole in our Galactic Center) are also briefly discussed, it was considered
 an opportunity  to evaluate a (tidal) charge of Reissner -- Nordstr\"om black hole
from observational estimates of shadow size in the Galactic Center and M87* done by the Event Horizon Telescope Collaboration based
its observations in April 2017.}


\begin{document}

\section{Introduction}

In May 2020, the gravity community celebrated the 80th anniversary of one of its brilliant representatives, Professor Remo Ruffini. The author had an excellent opportunity to talk about the enormous influence of Professor Remo Ruffini on the development of relativistic astrophysics in Russia.
For the first time, the author saw and followed Professor Remo Ruffini at the international conference organized in Leningrad in June 1988 (35 years ago) and dedicated to the 100th anniversary of the birth of Alexander Alexandrovich Friedmann, who 100 years ago, based on the analysis of solutions to Einstein's equations, concluded that the Universe should be evolving.
 Despite the gigantic significance of this theoretical discovery, in the USSR, the interpretation of Friedman's decisions to describe the behavior of the Universe was actually banned, since the model of the Universe where there is a beginning was considered by Soviet philosophers and ideologists to be too close to the idea of the divine creation of the world, therefore only the model of the Universe infinite in time and space was acceptable to Soviet ideology. In 1930s -- 1940s  Soviet philosophers and biased scientists denied the scientific and practical significance of quantum mechanics and relativity theory. Only by the early 1960s, due to the growing importance of scientists and especially physicists in the USSR, scientists had the right to legally consider models of the Universe that had a "beginning", proposed by Friedman, Lemaitre, Gamow and other authors. However, ideological restrictions have had an extremely negative impact on the development of research in the field of relativistic astrophysics.

In spite of unfavorable circumstances for this branch of research which sometimes were in some countries, currently, we are witnessing tremendous success in research in the field of relativistic astrophysics, in particular, the Nobel Prizes in physics were awarded for outstanding research in this field in 2017, 2019 and 2020. Remarkable results in shadows recontructions for M87* and Sgr A*
were obtained by the Event Horizon Telescope (EHT) Collaboration using observations of these objects in April 2017.
The EHT Collaboration found constraints on black hole charges (including electric charge of Reissner  -- Nordstr\"om black hole). In our recent papers
we generalized these results for black holes with a tidal charge based on our analytical expressions for shadow radius as a function of charge.

We organized paper in the following way. In Section \ref{development} a development of relativistic astrophysics in Soviet Russia was presented in brief. In Section \ref{Lemaitre} we describe briefly early cosmological studies developed by G. Lemaitre and G. Gamow. In Section \ref{Khaikin} we remind works of S. E. Khaikin (who was the founder of Soviet observational radioastronomy) and
we describe a history of CMB discovery in Pulkovo and reasons why the Soviet astronomical community chose not to notice this discovery.
In Section \ref{Remo}
we outline a huge impact of Professor Remo Ruffini on development of relativistic astrophysics.
In Section \ref{Shadow} we remind great achievements of the Event Horizon Telescope Collaboration in observations of Sgn A* and M87* and shadow reconstructions around the
black holes in these objects.
In Section \ref{tidal} we outline constraints on parameters of non - Schwarzschild spherical symmetrical black holes and  our contribution in these studies.
In Section \ref{conclusions} we present our conclusions.

\begin{figure}[th!]
\begin{center}
\includegraphics[width=0.32\textwidth]{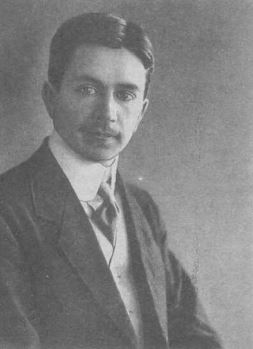}
\includegraphics[width=0.295\textwidth]{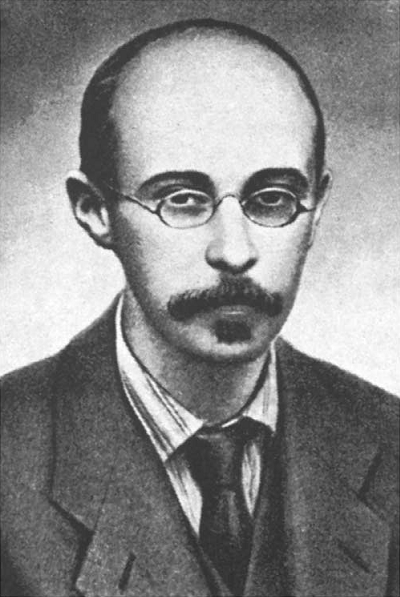}
\end{center}
\caption{V. K. Frederiks who was a founder of Russian schools in GR and theory of liquid crystals (left) and Alexander Friedmann who was the founder of mathematical  cosmology(right). }
 \label{Fig1}
\end{figure}

\begin{figure}[th!]
\begin{center}
\includegraphics[width=0.33\textwidth]{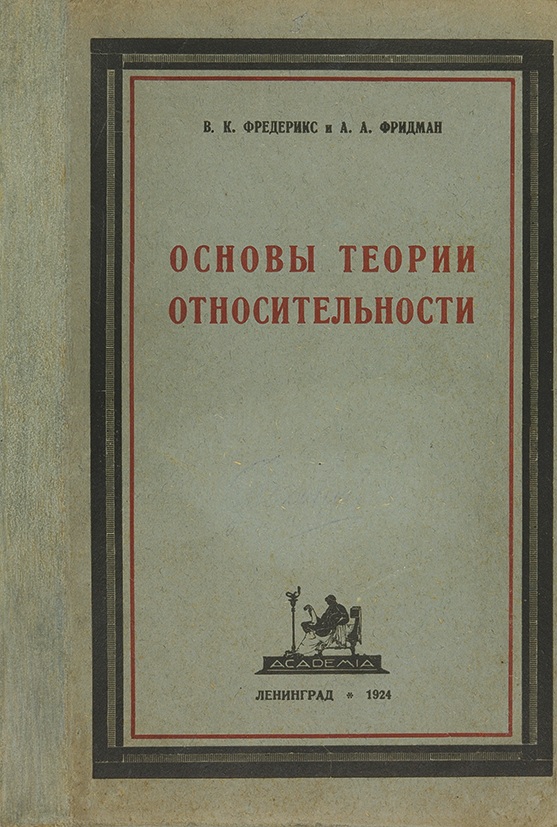}
\includegraphics[width=0.305\textwidth]{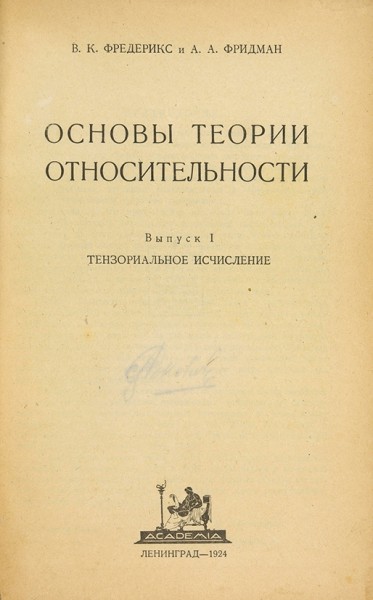}
\end{center}
\caption{Cover page (left) of joint book "Basics of General Relativity" by V. K. Frederiks and A. A. Friedmann and its first page (right). }
 \label{Fig2}
\end{figure}

\section{A development of relativistic astrophysics in Soviet Russia}
\label{development}

A development of general relativity in Russia started in 1918 after the end of WWI when Vsevolod Konstaninovich Frederiks came back from G\"ottingen where
he was an assistant of a famous mathematician David Hilbert (before that Frederiks was assistant with Professor Woldemar Voigt in G\"ottingen University).\footnote{During  WWI D. Hilbert paid a salary to a representative of an enemy country and many of Hilbert's colleagues did not support him in this action.}  It is well-known that at this period Hilbert was very interested in physics and specially in General Relativity.
In spite of great difficulties in times of the civil war in 1918 Soviet authorities permitted to establish
the main national physical journal "Physics -- Uspekhi" 	 (Advances in Physical Sciences)\footnote{\url{https://ufn.ru/en/.}},
the State Optical Institute under the leadership of D.S. Rozhdestvensky  and the Physical -Technical Institute (later it was named after A. F. Ioffe who was the founder and the first director of the Institution). Many famous scientists worked in these institutions, for instance,
four Nobel prize winners worked in the Ioffe Institute: N. N. Semenov, L. D. Landau, P. L. Kapitsa and Zh. I. Alferov. In Petrograd V. K. Frederiks
met Alexander Alexandrovich Friedmann\footnote{He was born on June 16, 1988 in Sankt-Petersburg.} and
convinced him that general relativity is the most interesting branch of theoretical physics and these two scientists decided to write  a  monograph on GR and the authors wrote only mathematical introduction in GR and this plan was not realized since  unfortunately, Friedmann died in 1925.
However, in 1924 V. K. Frederics and A. A. Friedmann published "Basics of general relativity" (Issue 1. Tensor calculus) in Russian (see, Fig.~\ref{Fig2}).

Before that, Friedmann found two non-stationary solutions of Einstein equations
\cite{Friedmann_22,Friedmann_24}.
Now we know that we live in the Friedmann world as the authors of the book \cite{Tropp_93} emphasized.

In spite of great achievements of Soviet scientists at the first stage of relativistic studies Soviet researchers
who are dealing with problems of quantum mechanics and relativity were under a strong press of official ideologists
since these people claimed these theories are based on idealistic philosophy and therefore, according to Lenin's philosophy doctrines
first, these theories are wrong (since correct physical theories must be based on dialectic materialism), second,
  research in the fields of these theories has only scholastic interest and has no value for the development of new industrial technologies.
  Western analysts of Soviet science came to similar conclusions \cite{Wetter_58,Graham_87,Kragh_12}.
  Ideological pressure on Soviet physicists involved in the atomic project has weakened since the mid-40s, since the creation of new technologies is incompatible with the presence of administrative restrictions. However, restrictions on discussions
of dynamic models of the Universe (proposed by Friedmann, Lemaitre, Gamow, etc.) existed until the early 60s of the last century and only about 60 years ago were gradually abolished.

\section{Early cosmological studies}
\label{Lemaitre}

\begin{figure}[th!]
\begin{center}
\includegraphics[width=0.35\textwidth]{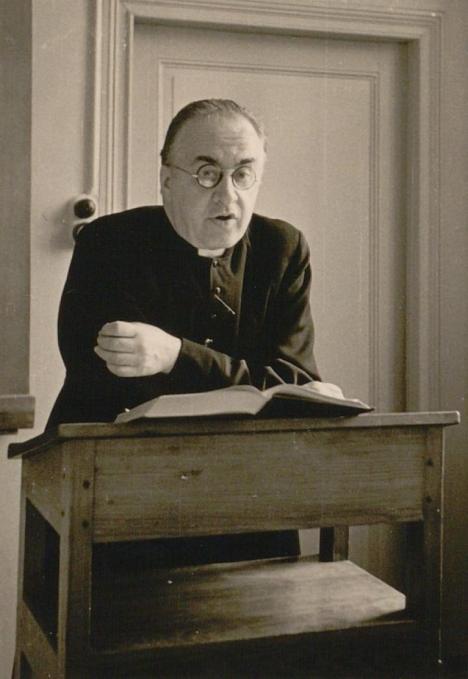}
\end{center}
\caption{Abb\'e
 Georges Lemaitre who firstly discussed  observational features of an Universe expansion and introduced a hot Universe model which was later called Big Bang. }
 \label{Fig4}
\end{figure}

Soon, after that George Lemaitre (see, Fig. \ref{Fig4}) considered a similar problem to earlier Friedmann's considerations and derived the redshift--distance relation which is called now as the
Hubble law \cite{Lemaitre_27,Lemaitre_31} which was found based on observational data \cite{Hubble_29}
 using redshifts for selected galaxies evaluated by Slipher \cite{Slipher_17}
 (interesting discussions of historical aspects of these discoveries are given in  \cite{Livio_11,Livio_13,Gron_18,ORai_19}).
Five years ago at the XXX General Assembly of the the International Astronomical Union in was proposed to call the expansion of the Universe
 as  the Hubble -- Lemaitre law (instead of the Hubble law)
 \cite{IAU_18}.

From these studies and corresponding documents it is known that Lemaitre intentionally deleted  his derivation of relation which is called now
the Hubble law from  paper \cite{Lemaitre_27} in its translation in \cite{Lemaitre_31}.
G. Lemaitre did not read Friedmann's papers as it was noted in \cite{Kragh_12a} and he had learnt about Friedmann's investigations only from conversations with A. Einstein at the Solvay conference in 1927.

 In contrast to Friedmann (who was a professional mathematician but he was a beginner in astronomy), Lemaitre was a skillful astronomer, he spent several years in the United States,
 worked under H. Shapley, defended his PhD at MIT in 1927 and he  had conversations with E. Hubble, V. Slipher  \cite{Kragh_18}.
Therefore,   Lemaitre knew remarkable observations of redshifts done by V. Slipher \cite{Slipher_17}\footnote{Due a great contribution
of Slipher in this field there ia proposal to add his name to characterize the Universe expansion and to call it as
 the Hubble -- Lemaitre -- Slipher law \cite{Elizalde_19}.} and he clearly understood that it was very important to find
 observational features of a proposed model and really the velocity -- distance could be such a test for a proposed cosmological model.
 Moreover, A. Eddington discussed radial velocities of spiral galaxies observed by V. Slipher as an important criterium to test a cosmological model
\cite{Eddington_24} (he considered the de Sitter model in  in his book).

In 1931, Lemaitre proposed a hot  Universe scenario \cite{Lemaitre_31_nature} which was later called the Big Bang model.
Really, J. Peebles called Lemaitre as the "Father of Big Bang Cosmology" in \cite{Peebles_71}.
Later, Lemaitre developed his idea concerning fireworks Universe model \cite{Lemaitre_34} where he discussed such an important feature of his model
as a background radiation with a liquid hydrogen temperature (or around several kelvins) and also in this paper Lemaitre interpreted the $\lambda$-term as a vacuum energy and later this idea was re-analyzed from a quantum field theory point of view by A. D. Sakharov in \cite{Sakharov_67}
(and later in Gliner's\footnote{A nice essay on Gliner's life,  his scientific works and general conditions in Soviet
 scientific community at this period
 were described in \cite{Yakovlev_23}.} paper \cite{Gliner_70}).
An extended essay on the fireworks Universe Model which was named by Lemaitre as Primeval-Atom
Universe is presented in \cite{Lemaitre_50}.
The {\it Big Bang} term ironically introduced by a famous British astronomer Fred Hoyle in 1949 (a genesis of this term  was discussed in \cite{Kragh_13}).

In January 1933 Lemaitre delivered his lecture "Primeval Atom Hypothesis" in Caltech.  A. Einstein followed the lecture
and noted  "it suggests too much the (theological) idea of creation"
after the lecture A. Einstein said "This is the most beautiful and satisfactory explanation of creation to which I have ever listened!"
\cite{Lambert}\footnote{The Soviet Union was an atheistic state and the discussion of church dogmas and their possible connection with natural phenomena were prohibited and the analysis of the models considered in the works of Friedmann and Lemaitre could be considered as an element of religious (and thus anti-state) propaganda.  In contrast to the "bourgeois" models of the Universe, in which there is a natural beginning of its evolution, Soviet philosophers and biased astronomers discussed models of the Universe, which is infinite in time and space. Now it can be seen that the ideological opposition of Soviet and bourgeois science had an extremely negative impact on the development of science in the USSR.}.

In \cite{Kragh_07,Luminet_14} Kragh,  Lambert and Luminet discussed important Lemaitre's insights proposed  in \cite{Lemaitre_34,Lemaitre_50}.

In 1946 G. Gamow proposed  a hot Universe model \cite{Gamow_46}, later in the framework of the approach he (with his co-authors) considered
chemical abundances of different elements \cite{Alpher_48}, while his colleagues Alpher and Herman evaluated a temperature of back ground radiation
which should be around $5^{\circ}$~K \cite{Alpher_48b}.
In his book Gamow gave a significantly higher temperature estimate around $50^{\circ}$~K \cite{Gamow_52} but later a bockground temperature
was  lower again around $6^{\circ}$~K \cite{Gamow_56}. In one of the last from his interviews Gamow noted that a few kelvins are consistent with his hot Universe model while several hundred kelvins are in a disagreement with the model \cite{Gamow_68}.
However,  at this period cosmology was not a fashionable field of research, for instance, an outstanding British astronomer Sir Martin Rees noted \cite{Rees_99} that "in 1950s cosmology was out of the mainstream of physics and only "eccentrics" like Gamow paid any attention to it".

When  the ideological ban for cosmological studies was lifted in the USSR Ya. Zeldovich considered a class of evolving Universe models including
a hot Universe model and he evaluated a background temperature as  $20^{\circ}$~K \cite{Zeldovich_64}.

\begin{figure}[th!]
\begin{center}
\includegraphics[width=0.32\textwidth]{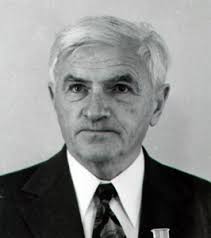}
\includegraphics[width=0.275\textwidth]{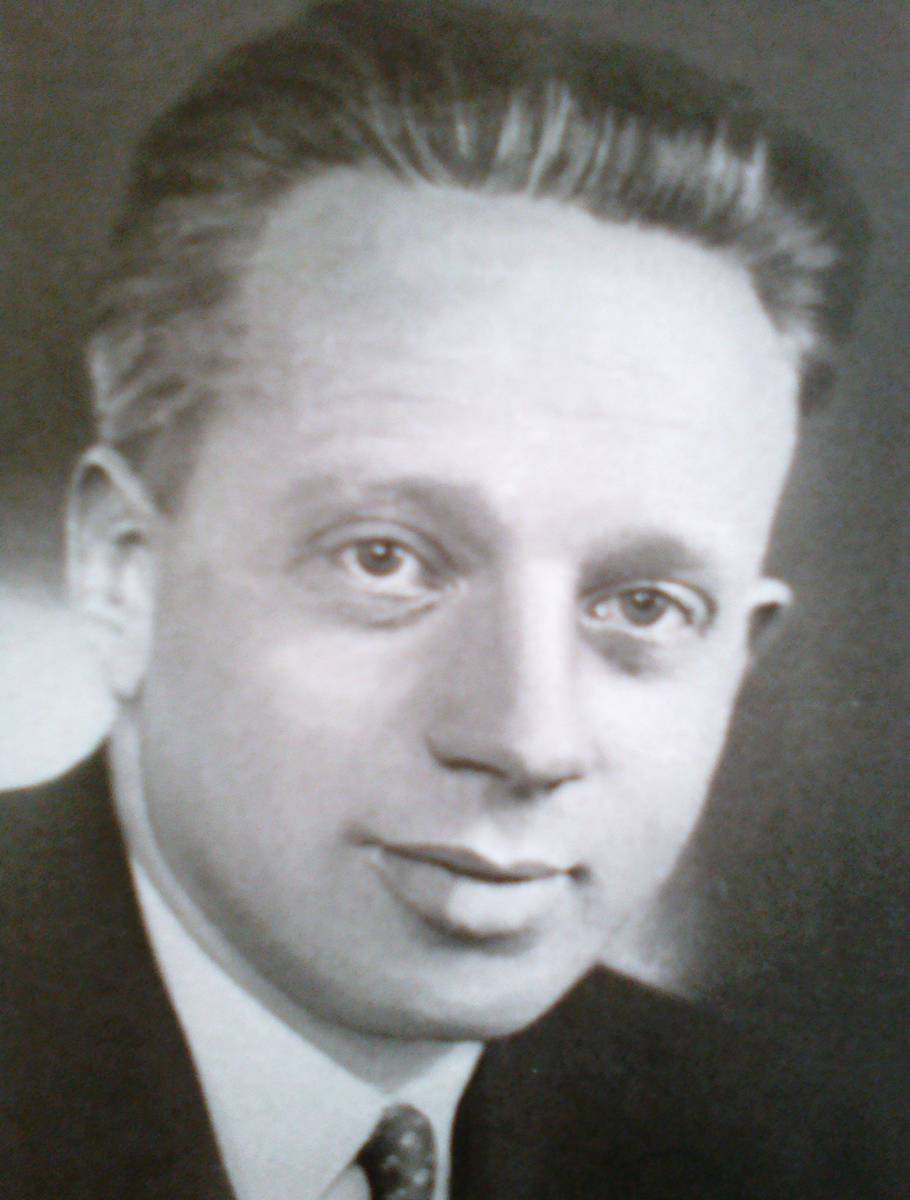}
\end{center}
\caption{Naum Lvovich Kaidanosky (left) and Semion Emmanuilovich Khaikin (right) who supervised T. A. Shmaonov at the  Pulkovo Observatory in 1950s. }
 \label{Fig3}
\end{figure}

\begin{figure}[th!]
\begin{center}
\includegraphics[width=0.32\textwidth]{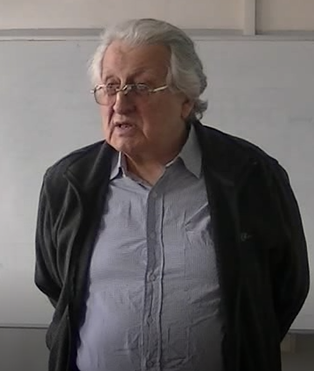}
\end{center}
\caption{Tigran Aramovich Shmaonov presents his talk about his discovery of CMB in 1957. The talk was delivered in the Institute for History of Natural Sciences and Technology in Moscow on 17 April 2017. }
 \label{Fig5}
\end{figure}

\section{S. E. Khaikin as one of the brightest representative of the Mandelstamm' school}
\label{Khaikin}

It is known that the relic radiation was discovered by T. A. Shmaonov at the Pulkovo Observatory in 1957 (before A. Penzias and R. Wilson), when Shmaonov was a PhD student of professors S. Khaikin and N. Kaidanovsky (see Fig. \ref{Fig3}).
The results were published in \cite{Shmaonov_57} (unfortunately, in 1957 papers published in this journal were not translated in English).
On 17 April 2017 in the Institute for History of Natural Sciences and Technology in Moscow T. Shmaonov delivered his talk about a history of the CMB discovery in 1957 (see,  Fig. \ref{Fig5}).
Many authors (including T. Shmaonov) claimed
 that neither in Shmaonov's entourage nor in the Soviet scientific astronomical and physical community did anyone know about the interpretation of this discovery as an evidence of the validity of the hot Universe model proposed by Gamow, see for instance, \cite{Trimble_06,Peebles_09,Cherepashchuk_13}.
  But this statement does not seem to me to be completely accurate (as it was firstly noted in \cite{Zakharov_20}). Firstly, the discussion of models in which the evolution of the universe is considered was actually banned, and secondly, G. Gamow, who left the USSR without the permission of the authorities, was considered a traitor and an enemy of the country, and it was extremely dangerous for a Soviet scientist to claim that Gamow's model was confirmed by observations. On the other hand, one of the Shmaonov' supervisors, S. E. Khaykin, as a representative of the Mandelstam school, was a widely educated physicist and it is difficult to imagine that Khaikin did not know Gamov's work on cosmology. In a few words, let's say about the works of Khaikin.

Leonid Isaakovich Mandelstam was a founder of an important school in Soviet Theoretical Physics and many Soviet physicists belong to this school,
including Igor Y. Tamm (who was a Nobel prize winner in physics in 1958 together with P. A. Cherenkov and I. M. Frank) and
Vitaly L. Ginzburg (who was a Tamm's student and a Nobel prize winner in physics in 2003 together with A. Abrikosov and  A. J. Leggett)\footnote{A nice essay about the Mandelstam's school (in particular, its discovery of combinational scattering in 1928) is presented in \cite{Fabelinskii_03}.}
Andrei D. Sakharov (who was also a Tamm's student, the father of the first Soviet thermonuclear bomb and a Nobel Peace Prize Laureate in 1975)
belongs to this school.
Mandelstam's range of scientific interests was very wide: from the theory of nonlinear oscillations, the basics of quantum mechanics and relativity theory to optics and radiophysics.
In 1925 Mandelstam came to Moscow from Leningrad to improve the level of teaching in theoretical physics and conducting scientific research in Moscow State University and his first students were A. A. Andronov, M. A. Leontovich and S. E. Khaikin.
In February 1928 G. S. Landsberg and L. I. Mandelstam discovered a combinational scattering of light one week before C. V. Raman and K. S. Krishnan
(now this phenomenon is called the Raman effect) \cite{Fabelinskii_03}
but Soviet researchers submitted their manuscript slightly later than Indian ones since during the period when it was necessary to prepare the article for publication, it turned out that Mandelstam's relative was arrested and sentenced, and only Mandelstam's petitions led to the replacement of the execution with a link \cite{Feinberg_02}.
In 1928 Mandelstam and Leontowitsch were among the first authors who considered tunnelling through a potential barrier analyzing properties of solutions of Schr\"odinger equation \cite{Mandelstam_28}. According to Tamm's remindings, Gamow was aware of these results, and they influenced his creation of the quantum mechanical theory of alpha decay in the same year. Gamow published his results in \cite{Gamow_28} (related studies on the subject are discussed in \cite{Stuewer_86}).

It is well-known that the first students of L. Mandelstam, namely, A. A. Andronov and A. M. Leontowich were outstanding representatives of Soviet science (Andronov was the founder of a school of non-linear physics in Nizhnij Novgorod, Leontowich was a leader of theoretical studies
in plasma physics and controlled thermonuclear fusion, while S. E. Khaikin (sometimes, in publications his family name was written
as Chaikin). However, he was an outstanding expert in physics in whole. In 1937 A. A. Andronov, A. A. Witt and S. E. Khaikin
published a fundamental book  "Theory of Oscillations" in Russian (all the authors were much younger than forty). When the book was published Alexander Adolfovich Witt was
accused of anti-Soviet activities, arrested and soon died in custody (he was subsequently rehabilitated due to the absence of corpus delicti).
Due to Witt's arrest, his surname was removed
from the list of authors of the first edition of the book (in subsequent editions his surname was restored).

Subsequently, the book was translated into English  \cite{Andronow_49} and it was edited by a famous American mathematician Solomon Lefschetz,
and in the preface to the English translation of this book, Lefschetz wrote that this is the first systematic presentation of research in the field of nonlinear oscillations in the world mathematical literature.
It should be noted that the translation of the book was carried out with the support of the research funds of the US Navy, which can be considered as a book having not only scientific, but also an important applied value. In 1960s an another English translation \cite{Andronov_66} and a German translation \cite{Andronov_65} were published. Thus, it can be argued that even after almost thirty years since the publication of the Russian edition of the book "theory of oscillations", the research presented in the book is of interest to physicists and mathematicians engaged in the study of nonlinear processes.

In 1930s Khaikin was the director of the Institute of Physics of Moscow State University and Dean of the Faculty of Physics, head of the Department of General Physics, lectured
on various sections of general physics, in particular, he wrote a university textbook on the course of mechanics. Unlike Soviet textbooks of those years, Khaikin's textbook did not contain references to the works of classics of Marxism - Leninism, at the same time, Mach's book "Mechanics" was recommended to readers, while Mach's works were not published in Soviet times, since his philosophy was criticized by Lenin.
In 1930s a number of researchers and professors from Leningrad and Moscow was arrested and significant part of them  was executed. In particular, Boris Mikhailovich Hessen, who was the director of the Institute of Physics of Moscow State University and the dean of the Faculty of Physics, as well as the deputy director of the Lebedev Physical Institute of the USSR Academy of Sciences, was arrested and soon shot. Following Hessen, Khaikin became the director of the Institute of Physics and Dean of the Faculty of Physics, that is, he took a position  that was potentially deadly at that time.
Khaikin was not only an outstanding researcher, but also a talented teacher, whose lectures were very much loved by students, however, some colleagues, inferior to Khaikin in their abilities as a scientist and teacher, wrote denunciations to the leadership of the University and
the party committee, which exercised ideological control not only over scientists and students, but also over the leadership of the university.

At the end of the Second World War, Khaikin's book "Mechanics" was being prepared for reissue. The new edition of the book received favorable reviews from prominent physicists, in particular, S. I. Vavilov and M. A. Leontovich, but Khaikin's detractors wrote to the university party committee complaining that the book develops machism and idealism in students and, in general, the book has a negative impact on Soviet students. In the end, Khaykin was dismissed from Moscow University, was the head of the department at the Moscow Mechanical Institute (which was created as part of the realisation of the Soviet atomic project and was subsequently named the Moscow Engineering Physics Institute) and Khaykin was also dismissed from this institute because of his political unreliability.  For several years in 1950s Khaykin worked at the Lebedev Physics Institute, but he had to leave it during the campaign against cosmopolitanism and cultural cringe.
In 1947 Khaikin headed an expedition of Soviet Academy of Sciences for astronomical observations during a solar eclipse in Brasil on May 20, 1947
and a group of radioastronomers (led by Khaikin) discovered emission from a Solar corona in radio band at the instant of full Solar eclipse
(two famous Soviet physicists V. L. Ginzburg and I. S. Shklovsky attended the expedition and later they explained radio emission (and other bands
of electromagnetic radiation) by synchrotron radiation in many astronomical sources including Solar corona).

In 1954, Khaikin left the Lebedev Physics Institute in Moscow and accepted an offer to organize a radio astronomy department at the Pulkovo (Main) Astronomical Observatory in Leningrad, and there he began creating new large radio telescopes, such as the large Pulkovo radio telescope and RATAN, which was built in the North Caucasus by his closest assistant Kaidanovsky.

Thus, in the 1950s (when Shmaonov discovered CMB) Khaikin could not declare a cosmological interpretation of this discovery, since the discussion of models of the evolving Universe was forbidden, a version of the Big Bang model was proposed by Gamow (who was persona non grata for Soviet science at that time) and it was extremely dangerous to claim that Khaikin's student confirmed Gamow's theory, personal stories of some employees of the Pulkovo Observatory who then we worked in Pulkovo and showed how cruel the reaction of the authorities can be to  unorthodox points of view on  purely scientific issues\footnote{In 1937 astronomer N. A. Kozyrev was convicted on trumped-up charges of anti-Soviet activities for 10 years of work in Siberian camps (subsequently, Kozyrev was fully rehabilitated). In custody, he received an additional sentence (10 years in the camps more)  for the following crimes: the defendant is a supporter of the idealistic theory of the expansion of the universe;
he stated that "being does not always determine consciousness";
he does not agree with the statement of Engels ("Dialectics of Nature") that "Newton is an inductive ass".
The main fault of the defendant is defined as "vulgarization of the teachings of K. Marx and F. Engels".
In 1950s as Khaikin, Kozyrev worked in Pulkovo Observatory. Kozyrev's biography was widely known in Soviet astronomical community}.

\begin{figure}[th!]
\begin{center}
\includegraphics[width=0.48\textwidth]{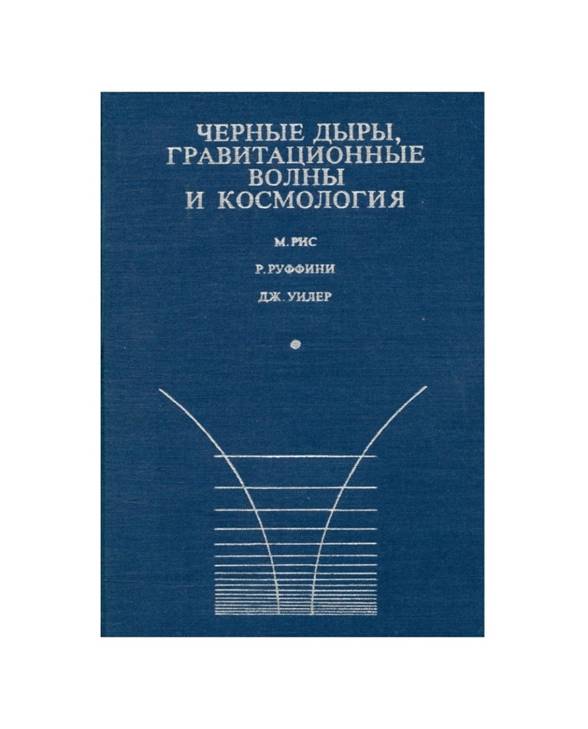}
\end{center}
\caption{Cover of Russian translation "Black holes, Gravitational waves and cosmology" by M. Rees, R. Ruffini and J. A. Wheeler. }
 \label{Fig6}
\end{figure}

\begin{figure}[th!]
\begin{center}
\includegraphics[width=0.26\textwidth]{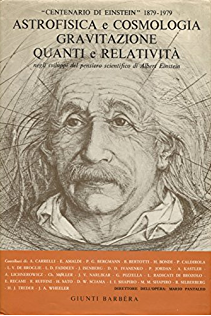}
\includegraphics[width=0.32\textwidth]{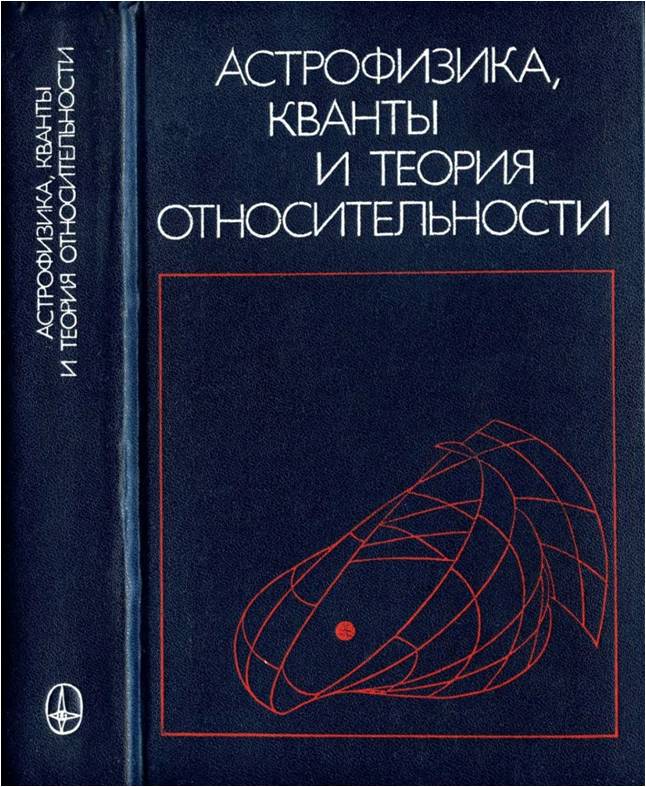}
\end{center}
\caption{Cover of the collection of articles "Astrofisica e cosmologia gravitazione quanti e relativita negli sviluppi del pensiero scientifico di Albert Einstein" (left panel) and its Russian translation (right panel).}
 \label{Fig7}
\end{figure}

\begin{figure}[th!]
\begin{center}
\includegraphics[width=0.5\textwidth]{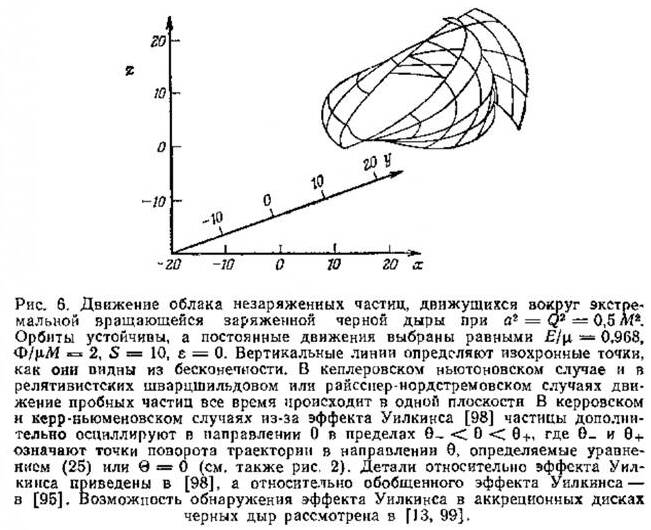}
\end{center}
\caption{Trajectories of uncharged particles in Kerr -- Newman metric from R. Ruffini's paper in \cite{Astrophysics_82} (results of corresponding calculations were presented in \cite{Johnson_74}.  }
 \label{Fig8}
\end{figure}

\begin{figure}[th!]
\begin{center}
\includegraphics[width=0.28\textwidth]{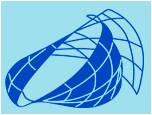}
\includegraphics[width=0.288\textwidth]{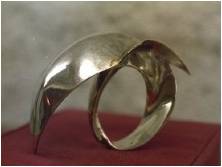}
\end{center}
\caption{The ICRANET logo (left) and the Marcell Grossmann prize (right).  }
 \label{Fig9}
\end{figure}

\begin{figure}[th!]
\begin{center}
\includegraphics[width=0.65\textwidth]{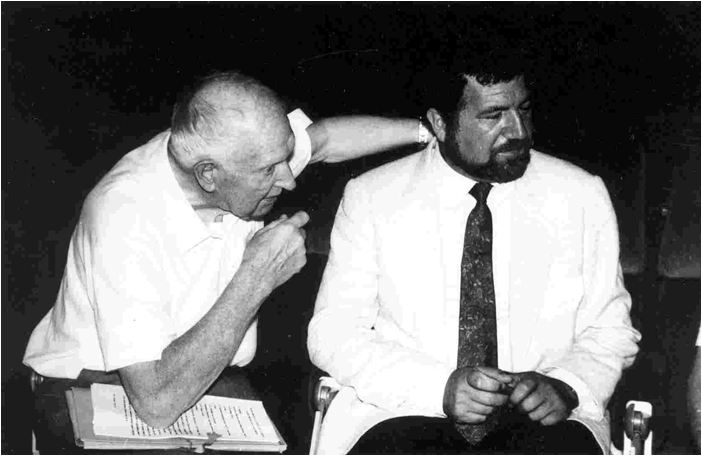}
\end{center}
\caption{Moisei Alexandrovich Markov (the Chairman of the Scientific Organizing Committee of the Friedmann-100 Conference and Academician Secretary of Nuclear Division of Soviet Academy of Sciences) and Professor Remo Ruffini at the Conference (Leningrad, 1988). }
 \label{Fig10}
\end{figure}

\begin{figure}[th!]
\begin{center}
\includegraphics[width=0.65\textwidth]{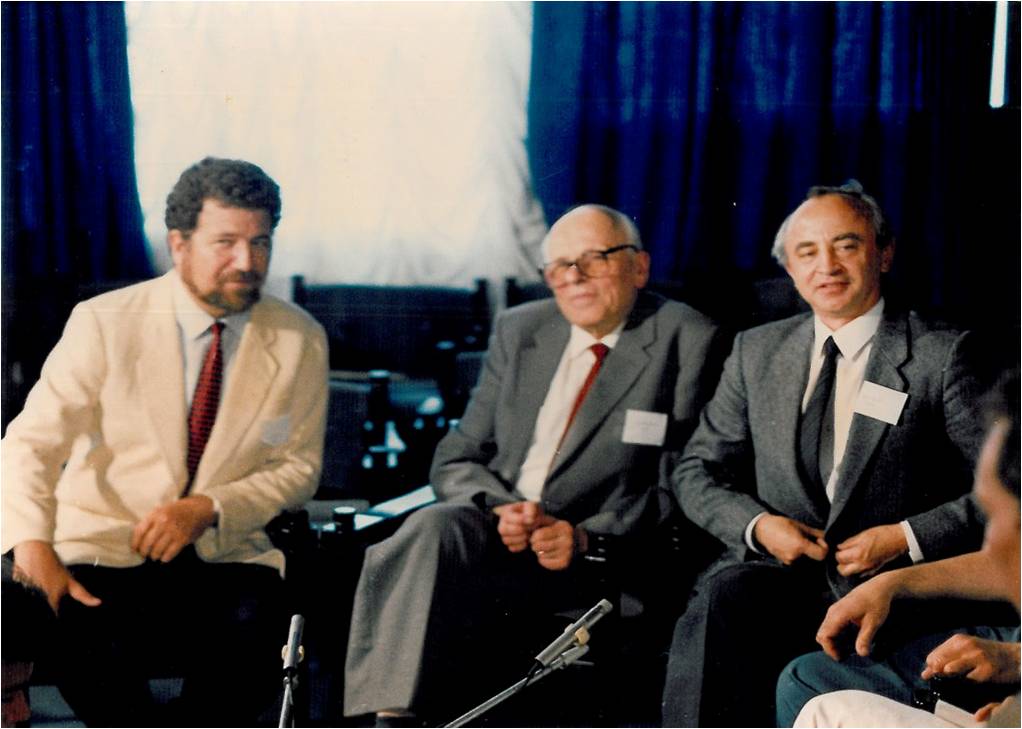}
\end{center}
\caption{Professor Remo Ruffini, Academician Andrei Dmitrievich Sakharov and professor Igor Dmitrievich Novikov at the Friedmann-100 Conference (Leningrad, 1988). }
 \label{Fig11}
\end{figure}

\section{A huge impact of Professor Remo Ruffini's studies on development of Relativistic Astrophysics}
\label{Remo}

In 1979 a scientific community celebrated a century since the birthday of A. Einstein and many good scientific books on general relativity and relativistic astrophysics have been published in the world and also in the Soviet Union. Many of these editions were translations from foreign languages.
Many of these scientific books were printed by Mir Publishing House. An advisory committee of this Publishing house consists of outstanding scientists
and they usually selected the best scientific books (especially in natural sciences) for translations in Russian.
Soviet reader knew that if a book was translated in Russian its authors are leading experts in the field since it was very careful choice
for translation of foreign scientific books.
For instance,
 In 1977 the Russian translation "Black holes, Gravitational waves and cosmology" (see, by M. Rees, R. Ruffini and J. A. Wheeler was published (translations of several books by J. A. Wheeler were printed in Soviet Union earlier).

In 1982 the Russian translation a collection of articles with title "Astrophysics, quanta and relativity theory" where papers of leading relativists were printed, including E. Amaldi,  H. Bondi, A. Lichnerowicz, C. Moller, R. Ruffini, J. A. Wheeler. I would like to note that instead of the Einstein portrait publishers selected the figure borrowed from Ruffini's paper where complicated trajectories of uncharged particles in a gravitational field of a Kerr--Newman black hole (Fig. \ref{Fig7}) and these trajectories are also shown in Fig. \ref{Fig8}. These trajectories are so nice and attractive
from artistic point of view, on the other hand, they are different from trajectories of test particles in point like source in Newtonian gravity, and later the orbits found in calculations done in \cite{Johnson_74} were used to create the ICRANet logo and the Marcell Grossmann prize as one can see in Fig. \ref{Fig8}.

An extraction of energy from Kerr black holes is an interesting astrophysical tasks and two phenomena have been proposed: namely Penrose process
 \cite{Penrose_71}  and Blandford -- Znajek  mechanism \cite{Blandford_77}. A new idea about an opportunity to extract energy from
 rotating black holes have been proposed in \cite{Rueda_22}.

Studies of classical books on GR (more correctly their Russian translations which available in bookstores and libraries) led
to apparence of papers \cite{Zakharov_86,Zakharov_91a,Zakharov_91b} where a qualitative and numerical analysis of geodesics in Kerr and Reissner -- Nordstr\"om metrics were presented.

\section{Shadow recontructions in M87* and Sgr A*}
\label{Shadow}

The  Event Horizon Telescope (EHT) Collaboration acts efficiently for several years. The  Collaboration uses a global network of telescopes acting
as a VLBI interferometer at 1.3 mm wavelength. Using EHT observations in April 2017 the EHT Collaboration reported about  shadow reconstructions in 2019 for M87* and in 2022 for Sgr A*. These remarkable pictures were reproduced elsewhere including newspapers and non-scientific journals.
As pioneer of VLBI technique L. I. Matveenko said more than 50 years ago, the EHT interferometer acts
as a telescope with Earth size.
In spite of huge differences in black hole masses and distances toward these objects (M87* and Sgr A*) the shadows have similar sizes (52$\mu as$ for Sgr A* and 42$\mu as$ for M87*). General relativity predicts size and shape of shadows around black holes but astronomers could observe
only bright structures and fortunately, there is synchrotron radiation which illuminate shadows and gives an opportunity to reconstruct their
size and shape.
However, the problem is very hard since an angular resolution of the interferometer (around 25~$\mu as$) is comparable with shadow sizes in M87* and Sgr A*.
These remarkable EHT achievements are based on three pillars: synchrotron emission which is generating in many astronomical objects including environments of supermassive black holes, VLBI ideas which were efficiently implemented in the EHT network and relativistic analysis of geodesics in the black hole metrics. In addition, to compare observations and simulations we assume an existence of black holes in these objects (it looks as a certain and clear assumption) and we adopt a model for an accretion flow (it does not look very certain and well defined hypothesis).

\subsection{Synchrotron radiation}

Electromagnetic radiation caused by moving electrons in magnetic fields  (which is now called synchrotron one) was discussed in details in a fundamental book by  \cite{Schott_12} but at these times there were no opportunities to detect it in experiments or in astronomical observations.
 In 1940s physicists started to build accelerators and it became possible to detect synchrotron radiation at accelerators and when observing astronomical objects with radio telescopes.
   Synchrotron radiation was re-discovered by I. Pomeranchuk and his co-authors \cite{Pomeranchuk_40,Iwanenko_44,Artsimovich_45}.
Later,  the first detection
of X-ray radiation from accelerated electrons in the General Electric 70-MeV synchrotron was reported by \cite{Elder_47}.

Famous Soviet astrophysicist I. S. Shklovsky reminded  that in forties of the last century he followed a talk about a discovery of radio emission from Sun and he concluded that  the radio emission from Sun was generated due to a synchrotron effect (in 1947 I. S. Shklovsky and V. L. Ginzburg were  members of the Soviet scientifc expedition in Brasil when Khaikin and his assistants discovered radio emission from solar corona).
He also concluded that the synchrotron effect is a cause of electromagnetic radiation in wide spectral band and this idea was the most brilliant from all his ideas in his entire scientific career as it was noted in his book  \cite{Shklovsky_91}.
For Crab Nebula Shklovsky interpreted electromagnetic radiation in wide spectral band (from radio to X-ray) as the synchrotron emission (\cite{Shklovsky_53,Shklovsky_76}).
Sir Martin Rees \cite{Rees_71} supposed that radio emission from extended radio source  may be explained by synchro-Compton radiation (in this case electrons are accelerating by electromagnetic waves).

Shklovsky was among the first authors who
 assumed that there is a black hole at the Galactic Center with mass around $3 \times 10^4 M_\odot$ \cite{Shklovsky_75}
and radiation has a non-thermal origin and probably a synchrotron radiation is responsible for a significant part of radiation from the Galactic Center (earlier, Linden-Bell and Rees emphasized arguments supporting a necessity for a presence of supermassive black hole at the Galactic Center with mass estimates in the range $[4\times 10^3, 10^7]M_\odot$  \cite{Linden_Bell_71}). In spite of the uncertainties in estimations of the black hole mass, ideas about a presence of a supermassive black hole at Sgr A* and the synchrotron emission from the Galactic Center region received confirmations in subsequent studies.

\subsection{Early VLBI in USSR}

Soviet radio engineer Leonid Ivanovich Matveenko was one of the first persons who understood an opportunity of inter-continental radio observations
and early history of these studies is described in paper \cite{Matveenko_07}.
In fall 1962 Matveenko reported ideas of VLBI in Pushchino at a seminar of
the Radio Astronomy Laboratory and he did not get a support to conduct such and experiment in Crimea as it was proposed.
However, these ideas were supported by participants of a seminar at Sternberg State Astronomical Institute (SSAI) of Moscow State University where
the SSAI director D. Ya. Martynov recommended to take out a patent due high scientific and technological importance of this proposal. Instead of patent a scientific paper on the issue has been published in Soviet journal "Radiophysics" \cite{Matveenko_65}. In this paper the authors proposed independent recording the signals and subsequent processing the data. In the initial version of the paper the authors proposed use a ground -- space interferometer but the editorial board of the journal recommended to remove this idea from the accepted version of the paper as it was noted by \cite{Matveenko_07},
In summer 1963  the director of the Jodrell Bank Observatory  B. Lovell visited Soviet Union as a guest of Soviet Academy of Science.   Matveenko delivered a talk about potential opportunities of interferometers with very large bases and Lovell noted that this idea looks feasible but he did not see any astronomical problem where such a resolution is needed \cite{Matveenko_07}. Both sides signed a memorandum on understanding about joint observations of Crimean and British telescopes at 32~cm wavelength. But these plans were not realized.


In the first paper on VLBI observations \cite{Matveenko_65} the authors discussed an opportunity ground -- space observations but these sentences were removed under request of the editorial board. However, later Soviet scientists and engineers formed a working group to develop a space radio antenna to act as a space component of a ground -- based interferometer. As Matveenko reminded the head of the project was V. P. Mishin, the scientific head was L. I. Matveenko, the chief engineer was V. I. Kostenko. It was assumed that the ground -- space interferometer will have an opportunity to observe compact maser sources and AGN at 1.35~cm wavelength.

In 1980s the idea on a Russian space -- ground interferometer (Radioastron) started to discuss again, but a preparation of the mission was very slow due to structural transformations in Russian economy.
It was expected that the interferometer would have an angular resolution at a level of a few microarcseconds at the shortest wavelength 1.3~cm as it was noted by \cite{Kardashev_88,Kardashev_01}. However, the space antenna was launched only in 2011 and Soviet astronomers lost an opportunity to built the first ground--space VLBI radio telescope and conduct observations in 1.35~cm wavelength with the best angular resolution before the realization of Japanese HALCA mission. The Radioastron mission was successfully launched in 2011 and was operating until 2019, scientific results after five years of operation are given in \cite{Kardashev_17}.

\subsection{Deflection of light and shadows around black holes}

In 1970s James Maxwell Bardeen  presented a picture of a dark region (a shadow) for  gedanken observations which correspond to a bright screen located
behind a Kerr black hole and a distant observer is located in the equatorial plane \cite{Bardeen_73}. Later,
Chandrasekhar
 reproduced a similar paper in his book \cite{Chandrasekhar_83}. However, neither Bardeen nor Chandrasekhar did not consider shadow as a possible test of GR since a) shadow sizes are extremely small to be detected for known black holes and b) there are no bright screens precisely behind black hole in astronomy. The authors represented a shadow shape as
a function $\beta(\alpha)$, where $\beta$ corresponds to impact parameter in rotation axis direction while $\alpha$ correspond to impact parameter in
in the equatorial direction.

Falcke et al. \cite{Falcke_00}; Melia and Falcke \cite{Melia_01}  simulated numerically a shadow formation for the Galactic Center in the framework of  a toy model, where the authors took into account electron scattering for
for a radiation in mm and cm bands. The authors concluded that  it is possible to observe  a dark region (shadow) around the black hole  in mm band, while it is not possible to see a shadow in cm band due to electron scattering. Consequent studies confirmed these conclusions.
In papers  \cite{Falcke_00,Melia_01} it was expressed an expectation to create a global network acting in 1.3~mm wavelength, therefore the best angular resolution of this interferometer is around  $25~\mu as$ (similar to the resolution of EHT network \cite{Akiyama_19}, while the shadow diameter was estimated as small as $30~\mu as$  assuming that the black hole mass is $2.6 \times 10^6M_\odot$ as it was evaluated in \cite{Eckart_96,Ghez_98}, therefore, expectations for shadow observations with these facilities were not very optimistic, however, now we know that the black hole mass is more $4 \times 10^6M_\odot$ and the EHT Collaboration reconstructed the shadow at Sgr A*.

In 2000s when the Radioastron mission was preparing for its launch it was expected that its the best angular resolution  was around 8~$\mu as$ at the shortest wavelength 1.3~cm and this angular value was comparable with the Schwarzschild diameter for the black hole at the Galactic Center since its mass was evaluated as high as $5 \times 10^6 M_\odot$   (based on black hole mass estimates done in \cite{Rees_82}). Therefore it was expected that observations with so accurate angular resolution will give an opportunity to find signatures of general relativistic effects. In papers in papers \cite{Zakharov_05a,Zakharov_05b} Zakharov et al. proposed to use  shadow observations around the Galactic Center as a test of presence of a supermassive black hole at Sgr A* since in the case of black hole mass around  $4 \times 10^6 M_\odot$ and a distance around 8~kpc toward the Galactic Center the shadow size is around $50~\mu as$.
Usually, there are no bright screens behind astrophysical black holes, however, following ideas proposed in \cite{Holz_02}, in paper \cite{Zakharov_05a} it was noted that a shadow should be surrounded by secondary images of many astrophysical sources and a presence of these secondary sources gives an opportunity to observe a shadow. It is important to note that a presence of a shadow depend only black hole metric and
it does not depend on uncertainties of our knowledge about accretion flows and only in the case if emitting regions are very close to
black hole horizons for rapidly rotating black holes the shadow sizes and shapes may be different from the standard case of bright screen behind a black hole.
In \cite{Zakharov_05a} it was  shown that in the case of an equatorial plane position of a distant observer the maximal impact parameter in the rotational axis direction is always (independently on $a$) $\beta_{max}(\alpha_{max})=3 \sqrt{3}$
while $(\alpha_{max})=2a$.
This claim was based on an analysis of critical curve for Chandrasekhar parameters $\eta (\xi)$ which separates scatter and capture of photons in Kerr metric. This analysis was done earlier in \cite{Zakharov_86}, see Fig. 2 in the paper and discussion therein and  if one considers critical values corresponding to multiple roots of the polynomial describing a radial photon motion as functions of  radial coordinate $r$  ($\xi(r)$, $\eta(r)$)and one has a maximal  $(\eta(\xi))$ at $\xi=2a$  one has  $\eta(2a) =27$ and $r(2a)=3$ (see also the critical curve Fig. 34 in page 352 in book
\cite{Chandrasekhar_83}). If $\eta(\xi)$ is known, one could obtain $\beta(\alpha$). Therefore, the function  $\eta(\xi)$  determines information about shadows for any position angle.

 Zakharov et al.  \cite{Zakharov_05a} expressed a hope that the shadow may be detected  if electron scattering may be ignored, in addition, the authors expressed a strong belief that the shadow can be detected with  VLBI network acting in mm band or with the projected ground--space interferometer Millimetron. The recent  results obtained by EHT Collaboration \cite{Akiyama_22a} where the shadow was reconstructed for Sgr A* remarkably confirmed our predictions. Earlier, a shadow was reconstructed for M87* \cite{Akiyama_19}. Later, there were presented polarization maps for M87* by \cite{Akiyama_21_a}
and possible distributions of magnetic fields were also given in  \cite{Akiyama_21_b} (polarization is connected with
synchrotron radiation of electrons accelerating in magnetic fields near M87*).

Based on the results of shadow size estimates for M87* done the EHT Collaboration  constrained charges of several metrics including Reissner -- Nordstr\"om,
Frolov, Kazakov -- Solodukhin and several other ones  \cite{Kocherlakota_21}. We would like to note that blue dotted line in the left panel Fig. 2 shown in  \cite{Kocherlakota_21} corresponds to an analytical expression for the shadow size as a function of charge done in
 \cite{Zakharov_14}.

\section{Shadows for Reissner -- Nordstr\"om black holes with a tidal charge}
\label{tidal}

A cosmic plasma is quasi-neutral it is  natural to expect that astrophysical black hole has a very small electric charge. In spite of these expectations we derived
an analytical expression for a shadow size as a function of charge \cite{Zakharov_05b} (we followed an approach used earlier in
\cite{Zakharov_91a,Zakharov_94}).  We also should to note that Reissner -- Nordstr\"om metric is a solution in the Randall -- Sundrum gravity theory with an extra dimension \cite{Dadhich_00}. Really, this solution looks like  Reissner -- Nordstr\"om metric but it is a generalization of this solution since parameter $q^2$ may be negative ($q$ is a black hole charge) and Dadhich et al. called it a Reissner -- Nordstr\"om metric with a tidal charge since
this additional parameter was caused by an existence of an extra dimension \cite{Dadhich_00}.
Later, it was proposed to adopt a Reissner -- Nordstr\"om metric with a tidal charge for the GC \cite{Bin_Nun_10}, however, it was shown that
a significant negative tidal charge is inconsistent with current estimates of a shadow size in Sgr A* since in this case a shadow size is much larger than its observed value \cite{Zakharov_12}.

Earlier we found allowed intervals for tidal charges based on EHT estimates of shadow sizes in M87* \cite{Akiyama_19} and Sgr A* \cite{Akiyama_22a}.
We will remind expression for
a Reissner -- Nordstr\"om black hole with a tidal charge in natural units ($G=c=1$) in  a form
\begin {equation}
  ds^{2}=-\left(1-\frac{2M}{r}+\frac{\mathcal{Q}^{2}}{r^{2}}\right)dt^{2}+\left(1-\frac{2M}{r}+\frac{\mathcal{Q}^{2}}{r^{2}}\right)^{-1}dr^{2}+
r^{2}(d{\theta}^{2}+{\sin}^{2}\theta d{\phi}^{2}),
\label{RN_0}
\end {equation}
where $M$ is a black hole mass, $\mathcal{Q}$ is its charge.
Constants $E$ and $L$ are connected with photon and they are describe photon geodesics, namely $E$ is photon's energy, $L$ is its angular momentum.
If we introduce normalized radial coordinate, impact parameter and charge
$\hat {r}=r/M, \xi=L/(ME)$,  $\hat{\mathcal{Q}}=\mathcal{Q}/M.$
We introduce also variables $l=\xi^{2}, q=\hat{\mathcal{Q}}^{2}$, then critical impact parameter corresponding to shadow radius \cite{Zakharov_14}
 \begin {eqnarray}
l_{\rm cr}=\frac{(8q^{2}-36q+27)+\sqrt{D}}{2(1-q)}, \label{RN_D_9b}
\end {eqnarray}
where
$D=-512\left(q-\dfrac{9}{8}\right)^3.$
As we noted earlier parameter $q$ may be negative for  a Reissner -- Nordstr\"om black hole with a tidal charge
(or for Horndeski scalar-tensor theory of gravity  \cite{Babichev_17,Zakharov_18b}).

The EHT Collaboration  evaluated the shadow radius in M87* and estimated parameters of several spherically symmetric metrics which
may be considered as alternatives for Schwarzschild metric in M87* \cite{Kocherlakota_21}.
In \cite{Zakharov_22} we generalizes results  \cite{Kocherlakota_21} for a Reissner -- Nordstr\"om black hole with a tidal charge  assuming
 similarly to \cite{Kocherlakota_21}, that angular diameter of a shadow in  M87* $\theta_{\text{sh~M87*}} \approx 3\sqrt{3}(1
\pm 0.17)\,\theta_{\text{g~M87*}}$, at confidence level around 68\% or $\theta_{\text{sh~M87*}} \in [4.31, 6.08] \theta_{\text{g}~M87*}$, where $\theta_{\text{g}~M87*} \approx 8.1~\mu as$, since $\theta_{\text{g~M87*}}=2M_{M87*}/D_{M87*}$ ($M_{M87*}=6.5 \times 10^9 M_\odot$ and $D_{M87*}=17$~Mpc, we found
$q \in [-1.22, 0.814]$ from Eq. (\ref{RN_D_9b}). In this case an upper limit for $q$ parameter ($q_{upp}=0.814$) corresponds to an upper parameter
   $\mathcal{Q}_{upp}=\sqrt{q_{upp}} \approx  0.902$, which corresponds to quantity calculated numerically and shown in Fig.~2 in \cite{Kocherlakota_21}.

Similarly to our previous estimates for tidal charge in  M87* in paper \cite{Zakharov_22c} we estimated a tidal charge for the black hole in GC.
We used estimates of shadow radius in GC from \cite{Akiyama_22a}. Following these studies, we assume that the shadow diameter in GC is
$\theta_{\text{sh~M87*}} \approx (51.8\pm 2.3)~\mu as$ at C. L. 68\% and in this case we obtain constraints for a tidal charge  $-0.27 < q <0.25$ at the same confidence level.

These results may be used for analytical estimates of charge for the Kazakov -- Solodukhin (KS) black hole.
Really,  Kazakov and Solodukhin considered a Schwarzschild black hole perturbed by quantum fluctuations \cite{Kazakov_94}.
 We should note that black hole with a negative tidal charge (or scalar-tensor charge in Horndeski gravity)
 could treated as a good approximation for KS black hole for a small KS charge, really
 according to Eq. (3.21) in  \cite{Kazakov_94} we have
\begin{equation}
  g(r)=-\frac{2M}{r}+\frac{1}{r}\left(r^2-{q_{KS}}^2\right)^{1/2} \approx 1-\frac{2M}{r}-\frac{{q_{KS}}^{2}}{r^{2}},
\label{KS_0}
\end{equation}
 where $q_{KS}$ is a KS charge. For small parameter $q_{KS}$  approximation
 we could use previous estimates for a KS charge in Sgr A* $(q_{KS})^2 < 0.27 $ ($(q_{KS}) < 0.52$).
  As we see in Fig. 2 in \cite{Kocherlakota_21} the shadow radius is growing as $q_{KS}$ is growing and
  it corresponds to the shadow diameter dependence of a tidal charge given in Eq. (\ref{RN_D_9b}).

\section{Conclusions}
\label{conclusions}

We recall a development of relativictic astrophysics in Soviet Russia and administrative constraints on development of relativistic astrophysics
in 1930s -- 1960s and this led to a significant lag in the field of research and as a result, the discovery of CMB in the research of Shmaonov turned out to be almost unnoticed by the world scientific community.
Russian translations of books by outstanding researchers such as Hawking, Wheeler, Penrose, Rees, Ruffini, Weinberg, Chandrasekhar contributed to activisations of studies in the field of relativistic astrophysics.
We remind contributions of Russian scientists in development of synchrotron radiation and its astrophysical applications to explain spectra of
astronomical objects. We also recall also the Matveenko's contribution in the development of the VLBI method for astronomical observations.
 Reconstructions of shadows
in M87* and Sgr A* give an opportunity to check GR predictions in these objects and to constrain parameters of alternative models for these
objects (including black hole charges) \cite{Akiyama_22c,Vagnozzi_22}.
Recent remarkable results of the EHT for reconstructions of shadows for black holes in Sgr A* and M87* showed a high efficiency of this technique.
We generalized these results for Reissner -- Nordstr\"om black hole with
a tidal charge (or a corresponding parameter in the Horndeski theory).
In papers \cite{Borka_13,Zakharov_16,Zakharov_18,Jovanovic_23} we described constraints for Yukawa theory parameter from observations bright stars, including observations  of the Schwarzschild precession for S2 star done by the GRAVITY collaboration \cite{GRAVITY_20}.


\subsubsection*{Acknowledgements}


The author also appreciates Professor Arthur Chernin and Professor Vladimir Soloviev for fruitful discussions of historical aspects of a Soviet cosmology development.

\end{paracol}

\end{document}